\documentclass[pdflatex,sn-mathphys-num]{sn-jnl}%

\usepackage{graphicx}%
\usepackage{multirow}%
\usepackage{amsmath,amssymb,amsfonts}%
\usepackage{amsthm}%
\usepackage{mathrsfs}%
\usepackage[title]{appendix}%
\usepackage{xcolor}%
\usepackage{textcomp}%
\usepackage{manyfoot}%
\usepackage{booktabs}%
\usepackage{algorithm}%
\usepackage{algorithmicx}%
\usepackage{algpseudocode}%
\usepackage{listings}%
\usepackage[left,mathlines]{lineno}
\usepackage{bm}
\usepackage{mathtools}

\theoremstyle{thmstyleone}%

\theoremstyle{thmstyletwo}%

\theoremstyle{thmstylethree}%
\newcommand{\CTS}{Co$_{1/3}$TaS$_2$} 

\begin{document}

\title[Article Title]{Spontaneous magnetic diode effect in \CTS}

\author*[1]{\fnm{Yining} \sur{Hu}}\email{yining.hu@mpsd.mpg.de}

\author[2]{\fnm{Taikang} \sur{Chen}}\email{taikang@pks.mpg.de}

\author[2]{\fnm{Jan} \sur{Priessnitz}}\email{priessnitz@pks.mpg.de}

\author[3,4]{\fnm{Zili} \sur{Feng}}\email{zlfeng@caltech.edu}

\author[3,4]{\fnm{Takashi} \sur{Kurumaji} }\email{kurumaji@caltech.edu}

\author*[3,4]{\fnm{Linda} \sur{Ye}}\email{lindaye@caltech.edu}

\author*[2,5,6]{\fnm{Libor} \sur{\v{S}mejkal}}\email{lsmejkal@pks.mpg.de}

\author*[1]{\fnm{Chunyu Mark} \sur{Guo}}\email{chunyu.guo@mpsd.mpg.de}

\author*[1]{\fnm{Philip J. W.} \sur{Moll}}\email{philip.moll@mpsd.mpg.de}

\affil[1]{\orgname{Max Planck Institute for the Structure and Dynamics of Matter}, \orgaddress{\city{22761 Hamburg}, \country{Germany}}}

\affil[2]{\orgname{Max Planck Institute for the Physics of Complex Systems}, \orgaddress{\city{01187 Dresden}, \country{Germany}}}

\affil[3]{\orgdiv{Division of Physics, Mathematics and Astronomy}, \orgname{California Institute of Technology}, \orgaddress{\city{Pasadena}, \state{CA} \postcode{91125},  \country{USA}}}

\affil[4]{\orgdiv{Institute of Quantum Information and Matter}, \orgname{ California Institute of Technology}, \orgaddress{\city{Pasadena}, \state{CA} \postcode{91125},  \country{USA}}}

\affil[5]{\orgname{Max Planck Institute for Chemical Physics of Solids}, \orgaddress{\city{01187 Dresden}, \country{Germany}}}

\affil[6]{\orgname{Institute of Physics, Academy of Sciences of the Czech Republic}, \orgaddress{\city{162 00 Praha 6}, \country{Czech Republic}}}

\abstract{
When electrons move through non-trivial spin textures in magnetic metals, conduction electrons couple to magnetic order through both their charge and their spin, acquiring non-trivial equations of motion and a dynamic spin polarization providing the foundation of spintronics. These mechanisms are, however, almost invariably reciprocal and the resistance a current encounters does not depend on its flow direction. Here we show that a chiral crystal lattice and an independently reversible chiral magnetic order together produce a diode whose forward direction is set by the magnetism rather than by the structure. In the triangular-lattice antiferromagnet \CTS, whose chiral lattice hosts a chiral magnetic order of its own, independent handedness, we observe a pronounced nonreciprocal resistance. 
Unlike conventional electrical magnetochiral anisotropy, which is driven by an external magnetic field, our spontaneous magnetic diode effect arises from an internal noncoplanar spin texture with vanishing net magnetization. Symmetry analysis and model calculations show that the effect originates from the interplay of noncoplanar spin order and spin-orbit coupling, which generates a chiral, time-reversal-symmetry-breaking electronic response. Crucially, reversing the magnetic handedness reverses the sign of the nonreciprocal resistance, demonstrating magnetic control of the diode polarity.
}

\let\originalprintabstract\printabstract
\renewcommand{\printabstract}{%
    \clearpage
    \originalprintabstract
}

\maketitle

\section*{Main}

The response of a conductor to an electrical current is ordinarily indifferent to the direction in which the current flows. Breaking this reciprocity has become a central theme in the study of quantum materials \cite{NagaosaYanase2024}, because a response that distinguishes forward from backward flow of signals and power yields a nonlinear, functional and possibly useful element as a diode. Such directional phenomena are broadly investigated, from superconducting diodes that pass a supercurrent one way but not the other \cite{Ando2020}, to unidirectional magnetoresistances in magnetic conductors and heterostructures \cite{avci2015unidirectional}, and quantum-metric driven non-linearity \cite{PhysRevLett.132.026301} as in MnBi$_2$Te$_4$~\cite{Wang2023}. Any such response that selects a direction of propagation can exist only where every symmetry that reverses that direction is broken, namely both time reversal on the one hand and spatial inversion and the mirror operations on the other \cite{Tokura2018}. Spatial inversion is straight-forwardly broken by system design via non-centrosymmetric crystal structures, asymmetric heterostructures or shapes. Time-reversal, when it comes to non-reciprocal transport, is usually broken by external magnetic fields. Akin to the anomalous Hall effect \cite{NagaosaAHE2010}, which may appear in zero field in materials with intrinsic time-reversal-symmetry breaking, one may find spontaneous non-reciprocity in magnetic conductors \cite{13pd-tlzp}. This anomalous non-reciprocity is almost overlooked as a response of bulk materials. Diodes of appreciable strength, and direction, set by an intrinsic magnetic degree of freedom would expose previously overlooked microscopic mechanisms of magneto-electric anisotropies and offer conceptual advantages in utilization as direct current non-linear readout becomes possible.

Here we report the observation of the \emph{spontaneous magnetic diode effect} (SMD), a strong spontaneous and switchable non-reciprocity in the non-collinear magnetic metal \CTS. This instrinsic response emerges from electronic motion in nontrivial magnetic textures (Fig. 1a) and represents an extra contribution to the nonlinear resistance \cite{PhysRevB.99.245153}:
\begin{equation}
R = R_0 + \alpha\,B^2 + \gamma_{ij}\,I_i B_j + \beta_{i}I_{i}.
\label{eq:resistivity}
\end{equation}

Here the second term is the well-known $B^2$ magnetoresistance. The third term emerges in noncentrosymmetric crystals which include chiral ones \cite{Tokura2018} and is historically referred to as \emph{electrical magnetochiral anisotropy} (eMChA) \cite{PhysRevLett.87.236602, PhysRevLett.95.256601, PhysRevLett.122.057206, PhysRevB.99.245153, Pop2014ElectricalMagnetochiral, Yokouchi2017ElectricalMagnetochiral}. The chiral crystals, where eMChA is typically observed, come in two mirror-image forms, or enantiomers, that carry opposite signs of $\gamma$ and therefore opposite diode polarity given a fixed relation between current and magnetic field. eMChA is observed in structurally chiral conductors such as tellurium \cite{PhysRevB.99.245153} and MnSi \cite{Yokouchi2017ElectricalMagnetochiral}, where it appears either as a field-induced directional asymmetry in the current–voltage characteristic or, equivalently, as a second-harmonic voltage under a low-frequency alternating current.
\begin{figure}
    \centering
    \includegraphics[width=1\linewidth]{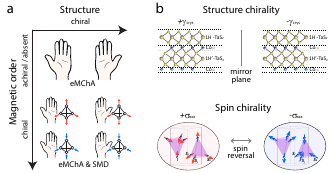}
    \caption{\textbf{a} Crystal structure and magnetic order as independent sources of the symmetry breaking required for \textit{electrical magnetochiral anisotropy} (eMChA) and \textit{spontaneous magnetic diode effect} (SMD). A chiral structure alone (left and right hands) gives rise to eMChA, which requires an external magnetic field. Chiral magnetic order, represented by all-out (red) or all-in (blue) spins on an upright tetrahedron, adds a second handedness. In \CTS, the chiral spin texture additionally enables an SMD at zero field. 
    \textbf{b} Top: two structural enantiomers of \CTS~related by mirror symmetry, labeled by the sign of their eMChA coefficient $\pm\gamma_\mathrm{crys}$. Bottom: spin textures with opposite scalar spin chirality in \CTS, related by spin reversal. They give second-order conductivities $\sigma_{xxx}$ of opposite sign.
    }
\end{figure}
In this work we introduce the last, fourth term which does not require external magnetic field, emerges due to internal symmetry breaking of the magnetic crystal and is thus spontaneous. Additionally, in our \CTS, it emerges even with vanishing net spin magnetization, contrasting it to dipole-field induced phenomena of ferromagnets. This SMD can therefore be viewed as an anomalous counterpart of eMChA, analogous to the relationship between the ordinary Hall effect and anomalous Hall responses arising from magnetic order or chiral spin correlations \cite{NagaosaAHE2010, Xie2026SpinNematic}.

\CTS~is a magnetic intercalation compound in which Co$^{2+}$ ions fill the van der Waals gaps of the dichalcogenide TaS$_2$ and form triangular layers of local moments. The ordered intercalation lowers the symmetry to the chiral, non-centrosymmetric space group $P6_322$ (No.~182) \cite{Park2022}, rendering it structurally chiral (Fig. 1b). The Co moments order in two stages. Below $T_{N1}\approx 38$~K a single-$Q$ stripe antiferromagnet breaks the threefold rotational symmetry and dominates the longitudinal resistance \cite{kirstein2025, feng2025}, while below $T_{N2}\approx 26.5$~K the moments adopt a non-coplanar arrangement that carries a finite scalar spin chirality (Fig. 1b) \cite{Park2023, takagi2023}. \CTS~has been intensely studied for its large anomalous Hall effect despite an almost vanishing net magnetisation of only $\sim 10^{-3}\,\mu_B$ per Co \cite{takagi2023}. These results have established \CTS~as a benchmark platform for chirality-driven topological transport and, as we argue here, an ideal host for the SMD.

\begin{figure}
    \centering
    \includegraphics[width=1\linewidth]{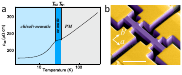}
    \caption{\textbf{a} Temperature dependence of longitudinal resistivity. \textbf{b} False-color SEM image of the device for transport measurement. The $b'$ axis denotes the crystallographic $b$ axis rotated by $30^\circ$ clockwise around $+c$ axis in device plane. The current is applied along $a$ direction. Scale bar 10 $\mu$m.}
\end{figure}

Detecting any non-linear transport requires high current densities difficult to attain in bulk crystals. We reach this regime by machining micron-scale bars (Fig. 2) from bulk single crystals with a focused ion beam (FIB), directing the current along the crystallographic $a$ axis through a cross-section of a few square microns (see methods for details). Miniaturization further avoids domain averaging as it scales the device below both the typical chiral and magnetic domain size, which do not coincide in \CTS~\cite{kirstein2025, kirstein2025topological}. The comparison between bulk and microstructured samples indicate that the fabrication preserves the intrinsic magnetism and introduces no detectable damage (Supplementary Figs.~S6). All results are fully reproduced in five devices, cut from three different single crystals. In all, the longitudinal resistivity confirms the established bulk behaviour, with a sharp cusp at $T_{N1}$ and a broader rollover at $T_{N2}$ (Supplementary Figs.~S1--S2). Equally, the magnetic phase diagram is unaltered in the microstructures compared to the bulk. These observations suggest that the micron-sized structures are representative of crystalline \CTS.

We measure two quantities simultaneously in a single device, the second-harmonic longitudinal voltage reflecting the diode behavior $V_{2\omega}$ and the Hall voltage $V_{b'a}$ (Fig.~3). The second-harmonic voltage traces a sharp hysteresis loop that mirrors the anomalous Hall effect, with flat plateaus between the switching fields, an abrupt sign reversal at the coercive field, and crucially a magnitude that remains finite down to zero applied field. The switching behavior of $V_{2\omega}$ excludes a self-heating origin of the non-linearity in agreement with a careful thermal analysis (see supplement 5). Indeed, the system remains a stable diode without an external field, with a state encoded by the magnetic polarity just like the anomalous Hall effect.

\begin{figure}
    \centering
    \includegraphics[width=1\linewidth]{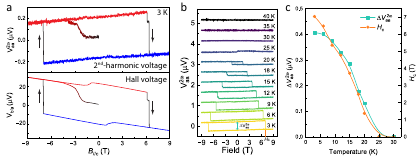}
    \caption{\textbf{a} The second-harmonic longitudinal voltage and the Hall voltage measured at 3 K. \textbf{b} Field dependence of second-harmonic longitudinal voltage at differerent temperatures, measured from 3 K till 40 K. \textbf{c} Temperature dependence of second-harmonic longitudinal voltage amplitude ($\Delta V^{2\omega}_{aa}$) and coercive field ($H_c$) extracted from \textbf{b}.}
    \label{fig_V2omega}
\end{figure}

After an initial cooldown in zero field, both $V_{2\omega}$ and the anomalous Hall effect are absent, the signature of a multidomain state in which the two magnetic states compensate. Polarizing the sample to about $5$~T selects a single magnetic state, after which the device switches reproducibly between the two polarizations on subsequent field cycles. The joint inversion of $V_{2\omega}$ and the anomalous Hall effect at the coercive field identifies the transition as a switch between the two spin-reversed non-coplanar textures, which reverses also the sign of the $\sigma_{xxx}$. The SMD response persists over a broad temperature range. Upon warming, the $V_{2\omega}(B)$ hysteresis loop progressively weakens and narrows (Fig.~3b), while both the hysteresis amplitude $\Delta V_{aa}^{2\omega}$ and coercive field $H_{\mathrm{c}}$ decrease and vanish at $T_{\mathrm{N2}}$ (Fig.~3c). Their common disappearance links the nonreciprocal response and its hysteretic switching to the low-temperature chiral magnetic order. 

That the measured second-harmonic voltage represents an anomalous, spin-geometry dependent magneto-chiral anisotropy, rather than a conventional one, follows from two independent observations. First, as the magnetic order itself is chiral, it could be a conventional eMChA driven by a real magnetic field in which the efficiency of the magnetic diode is enhanced. In this scenario, the magnetic order influences the sign and magnitude of $\gamma$. Indeed, chiral/helical spin textures are well-known to substantially enhance $\gamma$ as in CrNb$_3$S$_6$ \cite{PhysRevLett.122.057206}, which is structurally similar to \CTS~but hosts a distinct helical magnetic order, or as in MnSi \cite{Yokouchi2017ElectricalMagnetochiral}. Crucially, these interesting observations of enhanced yet conventional eMChA are fundamentally distinct from \CTS, as neither CrNb$_3$S$_6$ nor MnSi exhibits a zero-field diode effect. 

As \CTS~formally is a ferrimagnetic system with a finite magnetization along the $c$-direction, a second scenario may be conventional eMChA driven by the small but finite net magnetization, $V_{2\omega} \sim \gamma \mu_0(H+M) I$. While this would explain a finite signal at zero field and the hysteresis loop, also this scenario is clearly excluded by the data. $V_{2\omega}$ varies less than $20\%$ from zero field to $9~\mathrm{T}$, which would require an unphysically large magnetization of $\mu_0 M\simeq40~\mathrm{T}$ (see supplement 4). It is not surprising that the small zero-field magnetization of $0.009 \mu_B$ per Co ($\mu_0 M<10^{-3}~\mathrm{T}$) \cite{takagi2023}, which is far too weak to explain a linear response such as the anomalous Hall voltage, cannot drive a non-linear response \cite{Ghimire2018}. Uniquely here, the eMChA signal arises within a bulk material, distinct from recent observations in specially designed chiral ferromagnetic Ni nanotubes due to dipolar fields \cite{Xu2026}, nanosculpted helices of Co$_3$Sn$_2$S$_2$ due to chiral surface scattering \cite{Birch2026}, or uni-directional magnetoresistance in layered heterostructures due to designed symmetry breaking at the interfaces \cite{avci2015unidirectional}.

\begin{figure}
    \centering
    	\includegraphics[width = 0.8\linewidth]{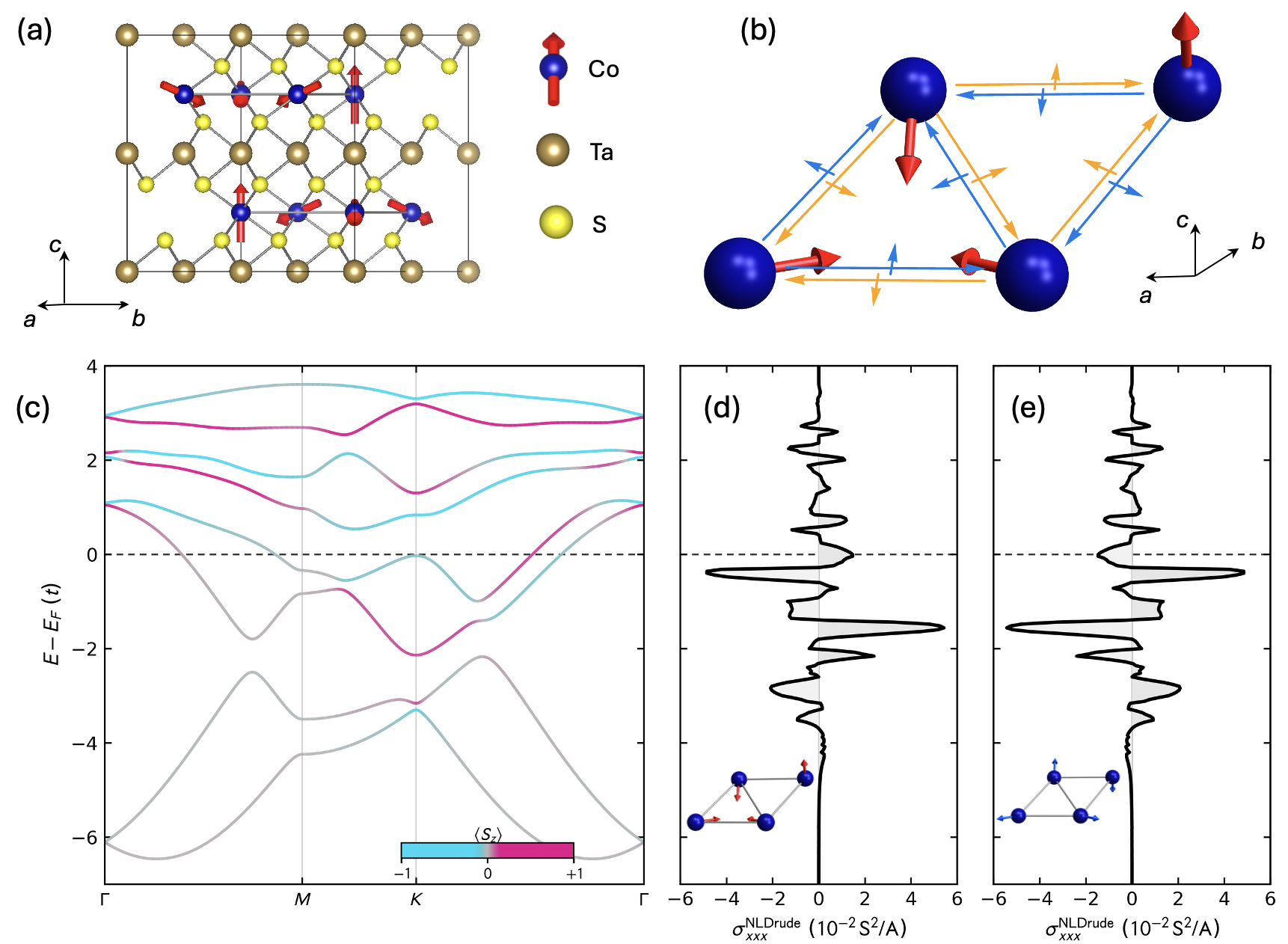}
   \caption{Theory of nonlinear conductivity in a model based on the \CTS. (a) Side view of the noncoplanar magnetic order in the \CTS~crystal. (b) Minimal model based on a single layer of the hexagonal Co lattice with marked spin-orbit coupling hoppings. (c) Model band structure calculated with spin-orbit coupling. The cyan and magenta colouring corresponds to computed spin polarization along the out-of-plane direction. (d) Calculated diode effect in terms of the Drude contribution to the nonlinear conductivity $\sigma_{aaa}^{NLDrude}$. (e)) The same as (d), but with reversed noncoplanar spin order.}
    \label{fig_theory}
\end{figure}

Next we discuss the microscopic origin of the large nonlinear conductivity in \CTS. Given its well-studied anomalous Hall effect with negligible net magnetization, it would be natural to assign the origin to the emergent topological field arising from the scalar spin chirality in this non-coplanar magnet~\cite{NagaosaAHE2010,takagi2023}. Surprisingly, however, our symmetry analysis excludes this topological origin and instead uncovers the necessity of spin-orbit coupling (SOC). To arrive at this conclusion, we consider the nonlinear conductivity corresponding to the SMD described by
\begin{equation}
    J_i = \sigma_{ijk}(\boldsymbol{S})\,E_jE_k,
\end{equation}
where $E_j$ is the applied electric field and $\sigma_{ijk}$ depends on the magnetic texture $\boldsymbol{S}$. The second-order conductivity $\sigma_{ijk}$ is related to the second-order resistivity $\beta_i$ in Eq.~(\ref{eq:resistivity}) via $\beta_{i} =  - R_0^3 \sigma_{iii}$. The allowed tensor components, both with and without SOC, were determined by analyzing the symmetry transformations of the nonlinear conductivity tensor under the corresponding spin and magnetic point groups (see supplement 7). The magnetically induced contribution to the nonlinear conductivity, transforming according to the Jahn symbol $V[V^2]$, contains the experimentally observed in-plane ($aa \equiv 11$ and corresponding to $\sigma_{aaa}$) response while forbidding the out-of-plane ($cc \equiv 33$ and corresponding to $\sigma_{ccc}$) component. This is indeed experimentally confirmed by the absence of $V^{2\omega}_{cc}$(see supplement Fig. S4). Yet crucially, the measured component $\mathbf{V}^{2\omega}_{11}$ is symmetry-forbidden in the absence of SOC, indicating that SOC is essential for the finite SMD.

From this symmetry analysis, we draw two key implications. First, the SMD requires SOC and is therefore unrelated to the nonrelativistic scalar spin chirality, pointing to a different mechanism beyond the topological one sensed by the anomalous Hall effect. Second, the necessary inversion symmetry breaking is generated spontaneously by the noncoplanar magnetic order. 
Our spontaneous diode effect is symmetry-distinct from the more conventional eMChA, which is described by a conductivity tensor transforming according to the Jahn symbol $eV[V^{2}]V$ and is induced by magnetic field. The two phenomena differ in three important respects. 
First, the eMChA contribution described by the tensor above reverses sign upon the magnetic field reversal, whereas the nonlinear conductivity considered here does not within the same hysteresis branch (Fig.~\ref{fig_V2omega}(a)).
Second, our effect is spontaneous, exhibiting a large remanent response even in the absence of an external magnetic field. Third, SMD switches sign under the reversal of the spin order, while the sign of the eMChA contribution does not depend on the internal spin order. 

To corroborate our symmetry analysis and these conclusions, we constructed a minimal tight-binding model of the full \CTS ~crystal, its side view shown in Fig.~\ref{fig_theory}(a). Figure~\ref{fig_theory}(b) shows the corresponding minimal model based on the Co hexagonal lattice with experimental noncoplanar magnetic order and imaginary SOC hoppings (see supplement 8 and Ref.~\cite{mitscherling2025orbitalmagnetizationparalleltransport} for the Hamiltonian). The SOC effectively emerges due to the broken in-plane mirror symmetry by the surrounding crystal structure as seen in Fig.~\ref{fig_theory}(a).

 Next we compute the Drude second-order conductivity $\sigma_{xxx}^{NLDrude}$ considered before in the context of unidirectional magnetoresistance~\cite{UnidirectionalNMS} (see supplement 8). When we neglect the SOC, the computed  $\sigma_{aaa}^{NLDrude}$ vanishes, corroborating our spin group symmetry analysis. Furthermore, we calculate  $\sigma_{aaa}^{NLDrude}$ with SOC switched on. We consider relatively large SOC ($\lambda$=0.5t, where $t$ is the kinectic electron hopping) due to the presence of heavy Ta around Fermi level in \CTS~as obtained in first-principles calculations\cite{Chen_Priessnitz_Smejkal_unpublished}. The strongly spin-polarized band structure is shown in Fig.~\ref{fig_theory}(c) and the calculated $\sigma_{aaa}^{NLDrude}$ corroborates four important aspects discussed above. First, it is nonzero only in the presence of SOC. Second, we obtain zero $\sigma_{ccc}^{NLDrude}$ in accordance with experiment and symmetry analysis. Third, when we consider  a relaxation time of 10~fs, we obtain large peaks near the Fermi level whose magnitude ($\sim$50~mS$^{2}$/A) is in the ballpark of the experimentally measured nonlinear conductivity ($\sim$300~mS$^{2}$/A). Finally, fourth, when we reverse the spins in our model, we obtain opposite sign of the $\sigma_{aaa}^{NLDrude}$ in perfect accordance with the aforementioned experimental observations and symmetry analysis. 

In the relatively clean regime, %
 the intraband Drude contribution is expected to dominate. Nevertheless, the more complex realistic electronic structure, characterized by the more complex spin texture and multiple band crossings in \CTS~\cite{Chen_Priessnitz_Smejkal_unpublished}, suggests that the nonlinear conductivity may be further enhanced through additional contributions such as the quantum metric dipole~\cite{Ulrich2026}. A full first-principles calculation of these effects is beyond the scope of the present work and is left for future investigation.

 Our nonreciprocal response is also distinct from the persistent nonlinear conduction predicted in $\mathcal{PT}$-symmetric antiferromagnets such as MnBi$_2$Te$_4$~\cite{Wang2023}. In particular, our calculated nonlinear Drude conductivity depends explicitly on the relaxation time, whereas the nonlinearity there originates from a quantum-geometric dipole and is a constant, independent of the lifetime.
 In contrast, \CTS~ belongs to a distinct symmetry class as a noncoplanar magnet, and its diode effect originates from the complex momentum-dependent spin texture. Consequently, the strength of the diode effect remains dependent on the carrier lifetime, as demonstrated by our calculations. In summary, instead of arising from an effective field associated with the scalar spin chirality, the strong SMD is related to a previously overlooked nontrivial spin texture and momentum-dependent spin splitting that break both time-reversal and spatial-inversion symmetries. These features emerge from the conspiration of noncoplanar spin order and spin-orbit coupling in \CTS.

The spontaneous non-reciprocity in \CTS~ presents an interesting window into the interaction of conduction electrons with structural chirality and complex spin textures.
Beyond the conceptual advance, we next discuss if new functionality can be generated from the phenomenon, as a first step towards utility. 

Like any field-driven nonreciprocity, the SMD is superimposed on a large resistive background. In a single crystal, however, crystallographic symmetry can be exploited to compensate the linear voltage via creating structures more complex than simple bars. We fabricated crystalline loops in an effective Wheatstone bridge configuration that suppresses the differential first-harmonic voltage, thereby increasing the relative prominence of the nonlinear response (Fig. 5a-b). This design utilizes the strong crystallographic anisotropy of \CTS~in two ways. First, because the longitudinal resistivities along $a'$ and $c$ are unequal, the bridge aspect ratio is fine-tuned such that the resistances of the $a'$- and $c$-oriented arms are matched at base temperature, leading to a near-vanishing linear response. 
 Second, we exploit the symmetry-dictated anisotropy of the SMD. Only the segments oriented along $a'$ generate SMD, whereas those along $c$ act as simple resistive bars with no second-harmonic response. 

\begin{figure}
    \centering
    \includegraphics[width=1\linewidth]{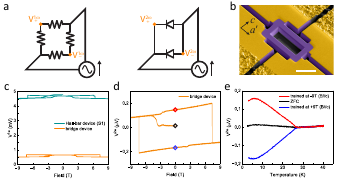}
    \caption{{a} Schematic of a Wheatstone bridge configured to cancel the linear-response voltage. The bridge is balanced at the first harmonic, $V_{+}^{1\omega}=V_{-}^{1\omega}$, but becomes unbalanced at the second harmonic owing to the anisotropic diode response, $V_{+}^{2\omega}\neq V_{-}^{2\omega}$. 
\textbf{b} False-colour SEM image of the bridge device. Scale bar, $10~\mu\mathrm{m}$. The $a'$ axis denotes the crystallographic $a$ axis rotated by $30^\circ$ clockwise around $+c$ axis.
\textbf{c} First-harmonic voltages measured in the Hall and bridge devices at $T=3$~K. 
\textbf{d} Second-harmonic voltage of the bridge device, measured simultaneously with the first-harmonic voltage in \textbf{c}. The $V^{2\omega}(B)$ hysteresis is comparable in amplitude to that of the Hall device in Fig.~3a, despite the strong suppression of the first-harmonic voltage. 
\textbf{e} Temperature dependence of the zero-field $V^{2\omega}$ after training at $\pm 9$~T with $B\parallel c$ and during zero-field cooling. The field-trained states produce opposite-sign responses that vanish at $T_{\mathrm{N2}}$, whereas the zero-field-cooled response remains close to zero.}
\end{figure}

 Experimentally, the bridge reduces the first-harmonic voltage by nearly an order of magnitude relative to that of a conventional Hall-bar device (Fig.~5c).
 In contrast, the second-harmonic hysteresis remains comparable in amplitude (Fig.~5d). 
 This demonstrates the size and robustness of the SMD, which allows assembly of crystalline elements into simple designed circuits with desired functionality.
 Further shape trimming could naturally adjust the linear and nonlinear signal level to any desired value.
Moreover, the near ideal $V_{2\omega}$ response function (fig. 5e) with a close-to-zero offset and a fully symmetric nonlinear signal demonstrates the accuracy of the FIB shape control.

The SMD should not be unique to \CTS, but it would exist more broadly in any non-trivial magnet with the required symmetry. These materials are chiral twice over, in both their lattice and spin texture, yet the two handednesses need not be locked. Where their coupling is weak, magnetic chirality can be reversed independently of the structural handness, making the sign of the diode a freely addressable variable rather than one fixed at growth. More broadly, our results demonstrate how more complex magnetic orders without internal dipole fields can act as sources of time reversal symmetry breaking which induces nonreciprocal transport.
For example, altermagnets \cite{PhysRevX.12.040501} may provide time-reversal symmetry breaking with vanishing net magnetization, as in Mn$_3$IrSi \cite{Hu2025, Onuki2022}. Whenever an emergent field can replace the role of applied fields, an effective field counterpart may be awaiting discovery.

Beyond its conceptual significance, the SMD offers a route to reading out information stored in anti-ferromagnetic orders, where conventional schemes based on net magnetization or stray fields fail. Here, the chirality of the magnetic state can already be read through the anomalous Hall effect, but this requires a dedicated set of transverse Hall contacts. The SMD, being a longitudinal response, can instead be detected in a simple two-terminal geometry.
More generally, the SMD effect joins a growing family of emergent nonreciprocal phenomena in quantum materials \cite{Tokura2018, NagaosaYanase2024}.
Recent developments in quantum materials have shifted from simple static responses (such as magnetization, electrical resistance, and dielectric permittivity ) to the rational design of dynamic response functions, for example the superconducting diode effect\cite{Ando2020, Ma2025, Moll2023}, unidirectional magnetoresistance \cite{avci2015unidirectional} and optical second-harmonic generation \cite{Toyoda2021, Tzschaschel2024}.
These generically rely on the spontaneous breaking of electronic symmetries and the appearance of typically forbidden responses. 
Here, a diode is produced by breaking of inversion and time-reversal symmetry, which will undoubtedly expose further exotic responses as other symmetry combinations are broken by nanoscopic control over lattice, spin and charge in the future.

\backmatter

\addcontentsline{toc}{section}{References}
\bibliography{sn-bibliography}

\section*{Methods}

\textbf{Crystal synthesis}\\
Single crystals of \CTS~were grown using a two-step procedure detailed in Ref. \cite{feng2025}. Polycrystalline \CTS~was first synthesized via solid-state reaction from stoichiometric mixtures of high-purity Co (99.99\%, Strem Chemicals), Ta (99.98\%, Strem Chemicals), and S (99.9995\%, Alfa Aesar) powders. The mixed powders were placed in an alumina crucible, sealed in an evacuated quartz ampoule, heated slowly to $900^\circ \mathrm{C}$, and held at that temperature for 24 hours.
Single crystals were subsequently obtained through chemical vapor transport using I$_2$ as the transport agent. The source and sink temperatures were maintained at $950^\circ \mathrm{C}$ and $850^\circ \mathrm{C}$, respectively. After approximately one week, plate-like crystals with lateral dimensions up to 2 mm were harvested. Phase purity was confirmed by powder X-ray diffraction.\\

\noindent\textbf{Device fabrication}\\
\CTS~microstructures were prepared using a Thermo Fisher Helios Ga FIB after the crystal orientation had been determined by single-crystal X-ray diffraction (Rigaku, analyzed with CrysAlisPro). Lamella for the five devices studied were extracted from the bulk crystal and transferred onto an araldite AB glue droplet on a prepatterned sapphire chip. To form good electrical contact between the lamella and the electrodes on the sapphire chip, the lamellae together with the sapphire chip were treated with radio-frequency (RF) Ar plasma etching for removing the native surface oxide, followed by high-power Au sputtering in the same chamber. After Au deposition, the lamella were patterned into the desired device geometries. \\

\noindent\textbf{Electrical measurements}\\
Electrical resistance was measured in a DynaCool PPMS using a multichannel SynkTek lock-in amplifier, with magnetic fields up to 9 T. For all devices, an AC excitation current from $400 ~\mathrm{\mu A}$ to $1470 ~\mathrm{\mu A}$ was applied by Keithley 6221 current source. Phase lock loop (PLL) is used to lock the demodulator to the excitation reference, so that it can track the frequency/phase precisely. The resistivity is read out by the first-order harmonic channel and the non-reciprocity is read out by the second-order harmonic channel. Across the full device set, clear MR and non-reciprocity were consistently observed, indicating that the FIB-fabricated microstructures retain their intrinsic transport properties. The FIB-induced amorphous surface layer, typically around $20~\mathrm{nm}$ thick \cite{Kato2004, Moll2018}, therefore represents a negligible fraction of the device cross-section (except for device S4) and does not affect the reported measurements.\\

\section*{Data availability}
The data supporting the findings of this study will be deposited on Zenodo.\\

\section*{Code availability}
The code supporting the findings of this study will be deposited on Zenodo.\\

\section*{Acknowledgements}
C. M. G. acknowledges financial support by the European Research Council (ERC) under grant Free-Kagome (Grant Agreement No. 101164280). The material synthesis efforts carried out at Caltech are supported by Gordon and Betty Moore Foundation through Moore Materials Synthesis Fellowship to L.Y. (GBMF12765).  Z.F. acknowledges support from Quantum Information and Matter (IQIM) Postdoctoral Fellowship at Caltech. We thank K. Wang for discussions related to the spin model calculations.\\

\section*{Author contribution}
Z.F., T.K. and L.Y. synthesized and characterized the crystals. Y.H., C.M.G. and P.J.W.M. designed the experiments, performed FIB microstructuring, conducted the transport measurements and analyzed the data. T.C., J.P. and L. \v{S}. developed the theoretical framework, including symmetry analysis, tight-binding modeling and calculations of the nonlinear Drude response. Y.H., C.M.G. and P.J.W.M. performed simulations of the spin model.  All authors contributed to writing the paper.

\section*{Competing interests}
The authors declare no competing interests.\\

\newpage\section*{Supplementary information}

\setcounter{figure}{0}
\renewcommand{\thefigure}{S\arabic{figure}}

\section{Scanning electron microscope images of all \CTS~bar devices}

\begin{figure}[H]
    \centering
    	\includegraphics[width = 0.98\linewidth]{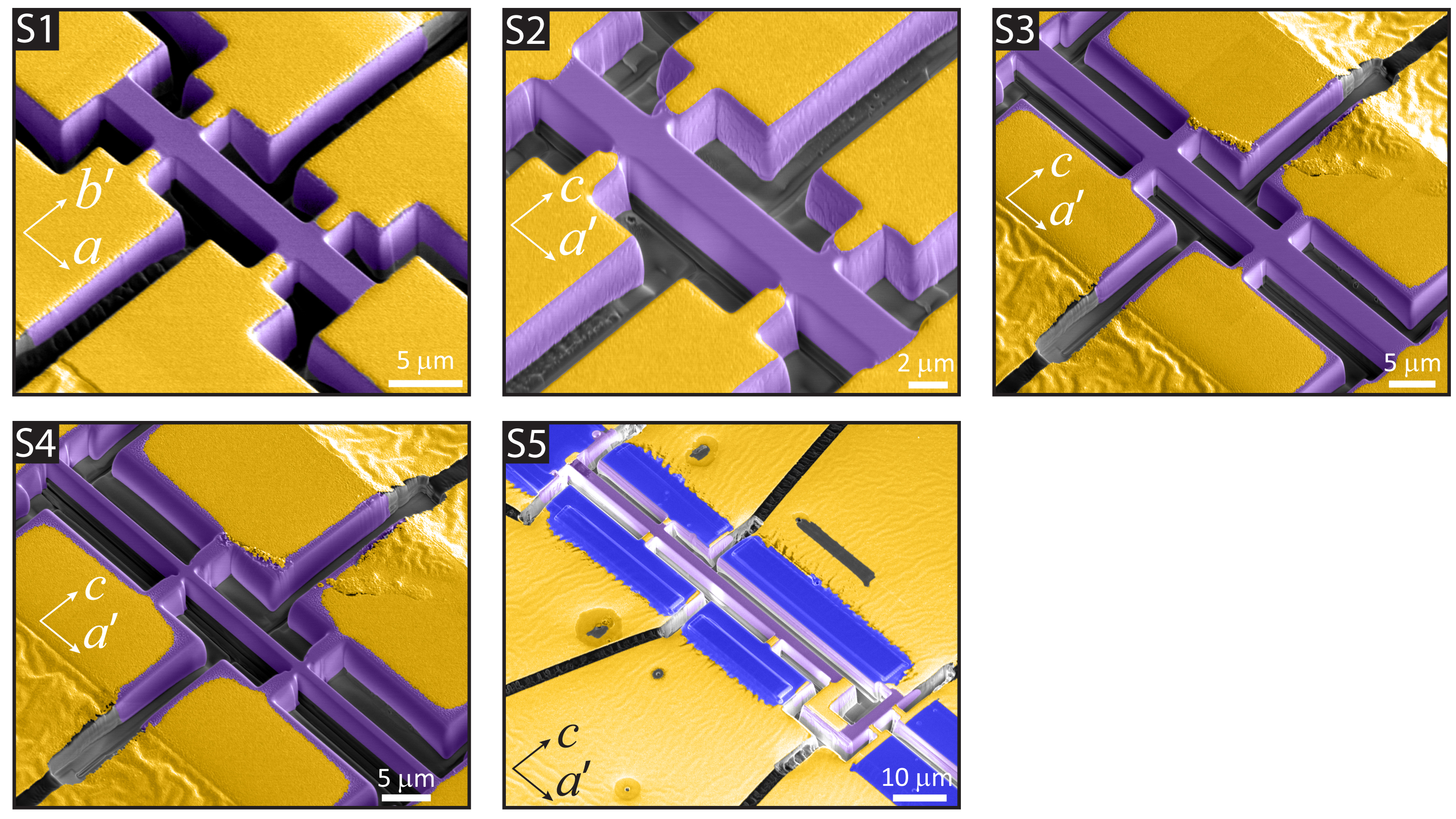}
    \caption{False-color SEM image of the devices. Au-coated contact pads are shown in yellow, the exposed \CTS~region in purple, and FIB-deposited Pt (blue) was used to secure the electrical contact. All microstructures were mounted on glue droplets to mitigate thermal-contraction mismatch and the resulting thermal differential strain at low temperature.}
    \label{figS1}
\end{figure}

Five microstructured devices were fabricated for transport measurements using the procedure described in the Methods, with a bar cross-section of approximately $\mathrm{3\times3 ~\mu m^2}$. Device S1 was patterned in the $ab$ plane; the current was applied along $a$ and the magnetic field along $c$, enabling measurements of the longitudinal response as well as the Hall voltage $V_{b'a}$. Devices S2-S5 were patterned in the $a'c$ plane; the current was applied along $a'$ and the magnetic field along $c$, and only longitudinal voltages were measured. Device S4 corresponds to device S3 after further thinning of the bar cross-section. In device S5, Pt film is to secure the electrical contact between the sample and the Au leads, by FIB-assisted deposition. Unless stated otherwise, the data in the main text are from device S1, data from the other devices are provided below.

\section{Temperature and field dependence of resistivity, second-harmonic longitudinal voltage and Hall voltage in all bar devices}
The resistivity of the five microstructures shows similar temperature dependence, with anomalies at $T_{\mathrm{N1}}\approx 38~\mathrm{K}$ and $T_{\mathrm{N2}}\approx 26.5~\mathrm{K}$, consistent with bulk \CTS~measurements \cite{Park2022}. Microstructure S4 is an exception: the slope changes at $T_{\mathrm{N1}}$ and $T_{\mathrm{N2}}$ are weaker, likely due to its reduced bar thickness, since it was fabricated by thinning S3, decreasing the bar thickness from $\mathrm{3.6~\mu m}$ to $\mathrm{1.4~\mu m}$. This geometric confinement can suppress long-range magnetic order and round the resistive kinks. Aside from this, the overall $\rho(T)$ dependence is consistent across devices, with differences mainly in the residual resistivity. This reproducibility indicates that the observed behaviour is intrinsically inherited from bulk.

\begin{figure}[H]
    \centering
    	\includegraphics[width = 0.5\linewidth]{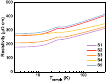}
    \caption{\textbf{Temperature dependence of resistivity in all devices.} Device S1 from crystal \#1, device S2-S4 from crystal \#2, device S5 from crystal \#3. }
    \label{figS2}
\end{figure}
All five devices show similar field dependence under the same measurement geometry (current along the crystalline $a$ axis; magnetic field applied along the $c$ axis).

For $T_{\mathrm{N1}} > T > T_{\mathrm{N2}}$, the magnetoresistance exhibits only the metamagnetic (nematicity) transition: the resistivity decreases smoothly as the $C_3$ rotational symmetry is restored. Below $T_{\mathrm{N2}}$, the transition shifts to lower fields due to coupling to the chiral order, and the magnetoresistance becomes hysteretic. The coercive field $H_c$ corresponds to the chirality-reversal field.

At a fixed temperature (e.g., 12~K), the nematic transition field $H_{\mathrm{nem}}$ is consistent across microstructures, lying slightly below 3~T. By contrast, $H_c$ varies substantially: at 12~K, $H_c \approx 4.5$~T in S1-S2, whereas it is already below 3~T in S5. Device S4 shows a strongly rounded chirality transition, consistent with geometric confinement.

The spread in $H_c$ likely reflects crystal-to-crystal differences, residual strain, and Joule heating. S5 was fabricated from a different single crystal, while S2 and S3 (from the same crystal) exhibit similar $H_c$, suggesting a crystal-dependent contribution. In addition, because the microstructures are mounted on glue rather than fully freestanding, differential thermal contraction between the glue and the microstructure can generate residual strain that shifts $H_c$. A later experiment of microstructured \CTS~mounted on free-standing membranes is worth to try for better understanding the strain effect on $H_c$.  Finally, Joule heating reduces the apparent $H_c$: for S5, increasing the excitation current from 0.3~mA (Fig.~S3) to 1.0~mA (Fig.~S4) lowers $H_c$ by $\sim 1$~T.

In all five devices, the non-nematic state exhibits a lower resistance than the nematic state. This trend is consistent with a preferential alignment of the stripes in the nematic domain along the long axis of the microbar ($Q$ oriented predominantly perpendicular to the bar)\cite{feng2025}. Two effects may contribute to this preference. First, although the Z$_3$ nematic domains are degenerate in an ideal bulk crystal, this degeneracy can be lifted in a sufficiently confined microbar, as a difference in the surface energy between the Z$_3$ nematic domains acts as an effective symmetry-breaking field that biases nematic domain selection. Second, a potential strain arising from differential thermal contraction between the microbar and the underlying glue during cooldown can also couple to the nematic order parameter.

\begin{figure}
    \centering
    	\includegraphics[width = 0.98\linewidth]{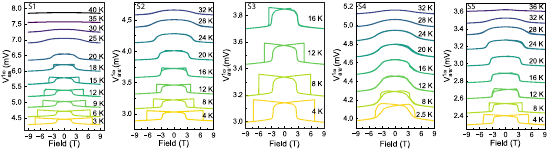}
    \caption{\textbf{Reproducibility of the magnetoresistance in all devices.} Data at different temperatures are vertically offset by 0.2 mV (S1), 0.15 mV (S2, S3) and 0.1 mV (S4, S5) for clarity. Current (density) for each measuremtn: $ \mathrm{1.4~mA~(2\times10^8A/m^2)}$ for S1, $ \mathrm{1.0~mA~(1\times10^8A/m^2)}$ for S2, $ \mathrm{1.0~mA~(9\times10^7A/m^2)}$ for S3, $ \mathrm{0.4~mA~(9\times10^7A/m^2)}$ for S4, $ \mathrm{0.3~mA~(3\times10^7A/m^2)}$ for S5.}
    \label{figS3}
\end{figure}

The second-harmonic signals measured in all five devices reproduce the key signature of SMD: a hysteretic loop emerges below $T_{\mathrm{N2}}$ as chiral order develops and widens upon cooling. The loop in S4 is again strongly rounded, consistent with geometric confinement. Device S5 exhibits an additional pair of non-antisymmetric kinks within the hysteretic loop when current is along $a'$, reducing the loop amplitude on the left side and enhancing it on the right. Because this feature lacks antisymmetry, we attribute it to an extrinsic origin, most likely the asymmetric device geometry of S5: the current enters along $a'$ direction and exits along $c$. This current path differs from that in S1--S4, for which the current path and associated heating profile are symmetric with respect to the measured bar. Device S5 doesn't show hysteresis loop in second-harmonic longitudinal voltage when current is along $c$.

\begin{figure}
    \centering
    	\includegraphics[width = 0.98\linewidth]{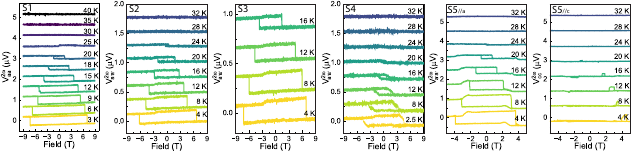}
    \caption{\textbf{Reproducibility of the SMD in all devices.} Data at different temperatures are vertically offset by 0.5 $\mu$V (S1), 0.25 $\mu$V (S2, S4),  0.3 $\mu$V (S3) and 0.9 $\mu$V (S5) for clarity. Current (density) for each measuremtn: $ \mathrm{1.4~mA~(2\times10^8A/m^2)}$ for S1, $ \mathrm{1.0~mA~(1\times10^8A/m^2)}$ for S2, $ \mathrm{1.0~mA~(9\times10^7A/m^2)}$ for S3, $ \mathrm{0.4~mA~(9\times10^7A/m^2)}$ for S4, $ \mathrm{1.0~mA~(9\times10^7A/m^2)}$ for S5//a, $ \mathrm{1.0~mA~(8\times10^7A/m^2)}$ for S5//c.}
    \label{figS4}
\end{figure}

$V_{b'a}$ could be measured only in device S1. It also onsets at $T_{\mathrm{N2}}$ and increases gradually upon cooling.

\begin{figure}[H]
    \centering
    	\includegraphics[width = 0.2\linewidth]{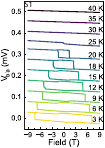}
    \caption{\textbf{Hall voltage of device S1 at different temperatures.} Data at different temperatures are vertically offset by 50 $\mu$V for clarity. }
    \label{figS5}
\end{figure}

\section{Preservation of bulk properties in microstructured devices}

\begin{figure}[H]
    \centering
    \includegraphics[width=\linewidth]{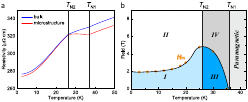}
    \caption{\textbf{Consistency of microstructured devices with bulk behaviour.}
    \textbf{a}, Temperature-dependent resistivity along the $a^{\prime}$ direction in microstructured device S5 (red) and its parent bulk crystal (blue). Vertical dotted lines mark $T_{\mathrm{N}2}$ and $T_{\mathrm{N}1}$.
    \textbf{b}, Field--temperature phase diagram of device S1. Orange circles denote the metamagnetic transition field $H_{\mathrm{m}}$ extracted from the magnetoresistance data in Supplementary Fig.~S3, and the black curve is a guide to the eye. Phases I--IV represent the four combinations of the presence or absence of non-coplanar chiral order and $C_{3z}$-symmetry-breaking order. The paramagnetic state occurs above $T_{\mathrm{N}1}$.}
    \label{fig:bulk_comparison}
\end{figure}

To assess whether focused-ion-beam processing alters the intrinsic magnetic properties of $\mathrm{Co}_{1/3}\mathrm{TaS}_{2}$, we compare the temperature-dependent resistivity of microstructured device S5 with that of its parent bulk crystal. The two traces have comparable magnitudes and temperature dependences, and exhibit anomalies at the same transition temperatures, $T_{\mathrm{N}1}$ and $T_{\mathrm{N}2}$ (Fig.~\ref{fig:bulk_comparison}a). The anomaly at $T_{\mathrm{N}1}$ is more pronounced in S5, consistent with reduced averaging over domains that break the threefold rotational symmetry ($C_{3z}$) in the smaller device. The dimensions used to convert the four-terminal resistance to resistivity are $2.48 \times 4.57 \times 33.07~\mu\mathrm{m}^{3}$ for S5 and approximately $469 \times 490 \times 45~\mu\mathrm{m}^{3}$ for the bulk crystal, corresponding to a difference in geometric volume of approximately $2.8 \times 10^{4}$. Within the resolution of these measurements, there is therefore no evidence that microfabrication shifts or suppresses the intrinsic magnetic transitions.

The field--temperature phase diagram obtained independently from device S1 provides a further comparison with the bulk behaviour (Fig.~\ref{fig:bulk_comparison}b). 
Below $T_{\mathrm{N}1}$, the metamagnetic boundary $H_{\mathrm{m}}(T)$, extracted from the magnetoresistance, separates phases with broken $C_{3z}$ rotational symmetry (phases I and III) from phases that preserve $C_{3z}$ (phases II and IV). The transition at $T_{\mathrm{N}2}$ separates phases with non-coplanar chiral order (phases I and II) from those without it (phases III and IV), whereas $T_{\mathrm{N}1}$ marks the boundary to the paramagnetic state. These boundaries therefore define four phases corresponding to the four possible combinations of chiral and rotational-symmetry-breaking order. The topology and characteristic temperature and field scales of this phase diagram are consistent with those reported for bulk single crystals in Ref.~\cite{kirstein2025}.

\section{Estimation of the effective field}

We define the normalized second-harmonic response as
\begin{equation}
    Y\equiv\frac{2V_{aa}^{2\omega}}{V_{aa}^{1\omega}}.
\end{equation}
For electrical magnetochiral anisotropy, this response is given by \cite{niu2023tunable}
\begin{equation}
    Y=\gamma I B,
\end{equation}
where $\gamma$ is the magnetochiral coefficient and $I$ is the applied current amplitude. In our case, $B$ is contributed by both the external field $B^{\mathrm{ext}}$ and the effective field $b^{\mathrm{eff}}$. If the two hysteretic states experience opposite effective fields, their responses can be written as
\begin{equation}
    Y_{\pm}
    =\gamma I\left(B^{\mathrm{ext}}\pm b^{\mathrm{eff}}\right).
\end{equation}

\begin{figure}
    \centering
    \includegraphics[width=0.75\linewidth]{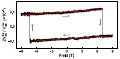}
    \caption{\textbf{Normalized second-harmonic response used to estimate the effective field.} Normalized response $2V_{aa}^{2\omega}/V_{aa}^{1\omega}$ of device S1 as a function of external magnetic field $B^{\mathrm{ext}}$ at $T=3~\mathrm{K}$. Grey arrows indicate the field-sweep directions. Red lines are independent linear fits to data acquired during the increasing- and decreasing-field sweeps before their respective switching transitions.}
\end{figure}

Here, $+$ and $-$ label the branches with positive and negative zero-field responses, respectively.
The linear portions of the two branches at $T=3~\mathrm{K}$ were fitted independently to
\begin{equation}
    Y_{\pm}(B^{\mathrm{ext}})
    =k_{\pm}B^{\mathrm{ext}}+Y_{0,\pm},
\end{equation}
where $k_{\pm}=\gamma I$ are the fitted slopes and $Y_{0,\pm}=\gamma I b^{\mathrm{eff}}$ are the corresponding zero-field intercepts. The fits yield
\begin{align}
    k_{+} &= 2.2710\times10^{-6}~\mathrm{T}^{-1},
    &
    Y_{0,+} &= 1.01562\times10^{-4}, \\
    k_{-} &= 2.3524\times10^{-6}~\mathrm{T}^{-1},
    &
    Y_{0,-} &= -8.3319\times10^{-5}.
\end{align}
Because the two slopes differ by only approximately $3.5\%$, we use their arithmetic mean,
\begin{equation}
    \overline{k}
    =\frac{k_{+}+k_{-}}{2}
    =2.312\times10^{-6}~\mathrm{T}^{-1}.
\end{equation}
The effective-field contribution reverses sign between the two branches. The corresponding effective intercept is therefore obtained as half of the difference between their zero-field intercepts:
\begin{equation}
    Y_{0,\mathrm{eff}}
    =\frac{Y_{0,+}-Y_{0,-}}{2}
    =9.244\times10^{-5}.
\end{equation}
The magnitude of the effective field is obtained by dividing this effective intercept by the average slope:
\begin{equation}
    \left|b^{\mathrm{eff}}\right|
    =\frac{Y_{0,\mathrm{eff}}}{\overline{k}}
    =\frac{9.244\times10^{-5}}
           {2.312\times10^{-6}~\mathrm{T}^{-1}}
    \simeq40~\mathrm{T}.
\end{equation}
This analysis therefore indicates an effective field $b^{\mathrm{eff}}$ on the scale of several tens of tesla. Because the extraction assumes a common linear response for the two branches, we regard $b^{\mathrm{eff}}\sim40~\mathrm{T}$ as an estimate of the characteristic field scale rather than a precise determination.

\section{Thermal effect on $V_\mathrm{2\omega}$}
\begin{figure}[H]
    \centering
    	\includegraphics[width = 0.98\linewidth]{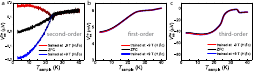}
   \caption{Temperature dependence of the \textbf{a} second harmonics of longitudinal response, \textbf{b} first harmonics of longitudinal response and \textbf{c} third harmonics longitudinal response (from device S1).}
    \label{figS6}
\end{figure}
Joule heating is a common extrinsic source of higher-harmonic voltage generation. An a.c. current $I(t)=I_{\omega}\sin(\omega t)$ produces a temperature component oscillating at 2$\omega$. Unequal resistances at the current contacts can cause asymmetric heating, generating an oscillating temperature difference between the voltages probes and hence an extrinsic $V_\mathrm{2\omega}$. Unlike the intrinsic non-reciprocity signal, the sign of which is flipped by the direction of the magnetic field applied, the extrinsic signal stays nearly field-independent as the thermal imbalance depends mainly on the device geometry rather than the switch of scalar spin chirality.

To quantify the temperature evolution of the thermal effect induced offset of second-harmonic voltage, we performed zero-field cooling and warming up after training in a $\pm$ 9T magnetic field in device S1. As shown in Fig. S8a, the offset remains positive over the measured temperature range. Consistent with a thermal origin, the third-harmonic signal follows a trend proportional to $R\cdot dR/dT$, indicating that Joule heating is non-negligible. The close agreement among the three third-harmonic traces indicates that Joule-heating-related artefacts do not produce hysteresis under our measurement conditions. Therefore, the hysteresis loop observed in the second-harmonic channel cannot be attributed to heating, and therefore reflects a nontrivial non-reciprocal transport response. 

\begin{figure}
    \centering
    	\includegraphics[width = 0.98\linewidth]{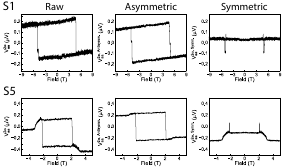}
   \caption{Decomposition of field-asymmetric and symmetric components of the second-harmonics at 12 K in device S1 and S5//a.}
    \label{figS7}
\end{figure}
This is further elaborated in the analysis of the field-asymmetric/symmetric component of $V_{2\omega}$ signal across different devices. To keep the loop feature in the asymmetric composition, we need to separate the data measured during the up-sweeping and down-sweeping of the field into $V_{\mathrm{sweepUP}}(B)$ and $V_{\mathrm{sweepDOWN}}(B)$, and do calculation as below:
\begin{equation}
    V_{\mathrm{asymm}}(B) = (V_{\mathrm{sweepUP}}(B)-V_{\mathrm{sweepDOWN}}(-B))/2;
\end{equation}
\begin{equation}
    V_{\mathrm{symm}}(B) = (V_{\mathrm{sweepUP}}(B)+V_{\mathrm{sweepDOWN}}(-B))/2.
\end{equation}

While the asymmetric component is nearly identical across devices, the symmetric component differs distinctively between devices S1 and S5. In device S1, it remains vanishingly small across the entire field range; the two sharp peaks at the critical field are most likely attributable to the symmetrizing analysis process rather than to intrinsic behavior. In device S5, however, the symmetric component is clearly observed and exhibits a distinct field dependence. This clear difference between these devices originates directly from their different geometry. The hall-bar-like structure in device S1 ensured a well-defined thermal boundary condition, thereby strongly reducing the thermal gradient across the microstructure. Comparatively, device S5 is shaped into an "L"-like geometry, in which thermal gradient is readily expected due to the asymmetry of the device itself and therefore generates a non-negligible field-symmetric $V_{2\omega}$ component.

\section{Comparison between $V_{2\omega}$ and $V_\mathrm{H}$}

Despite the overall similarity between $V_{2\omega}$ and the Hall signal, there are important differences between them (Fig. S10a). While the anomalous Hall effect acquires a clear feature at the meta-magnetic transition, $V_{2\omega}$ passes through it unmarked in the same device. The reason is that the Hall voltage subsumes all time-reversal odd effects in addition to the effective field, such as the normal Hall effect and an ordinary contribution proportional to the net magnetization. Naturally it therefore inherits a step when the magnetization varies, while the $V_{2\omega}$ responds to the effective field alone. The clean separation of the two signals thus establishes the $V_{2\omega}$ as a probe of the non-reciprocity distinct from the Hall voltage.

To understand how the non-coplanar spin texture evolves under a magnetic field and how this evolution affects the scalar spin chirality and the net magnetization, we construct a spin Hamiltonian and determine the lowest-energy spin configurations using a variational Ansatz for different magnetic fields.

The Hamiltonian is \cite{PhysRevMaterials.6.024405}
\begin{equation}
\begin{aligned}
\mathcal{H} =
&\;\frac{J_1}{2}\sum_{\langle i,j\rangle}
\mathbf{S}(\mathbf{r}_i)\cdot\mathbf{S}(\mathbf{r}_j)
+\frac{J_2}{2}\sum_{\langle\!\langle i,j\rangle\!\rangle}
\mathbf{S}(\mathbf{r}_i)\cdot\mathbf{S}(\mathbf{r}_j)
+ \sum_{\langle i,j\rangle}
\mathbf{D}_{ij}\cdot
\left[\mathbf{S}(\mathbf{r}_i)\times\mathbf{S}(\mathbf{r}_j)\right] \\
&+ \frac{B}{2}\sum_{\langle i,j\rangle}
\left[\mathbf{S}(\mathbf{r}_i)\cdot\mathbf{S}(\mathbf{r}_j)\right]^2
+ K\sum_i \left(S_i^{z}\right)^2
- \sum_i H_\mathbf{ext}\cdot\mathbf{S}(\mathbf{r}_i).
\end{aligned}
\end{equation}

where $J_1$ denotes the nearest-neighbor in-plane exchange interaction and sets the overall energy scale, $J_2$ is the next-nearest-neighbor in-plane exchange interaction, $D$ characterizes the strength of the Dzyaloshinskii--Moriya interaction, $B$ is the biquadratic exchange strength, $K$ represents the uniaxial single-ion anisotropy, and $\mathbf{H}_{\mathrm{ext}}$ is the applied external magnetic field.

\begin{figure}
    \centering
    \includegraphics[width=1\linewidth]{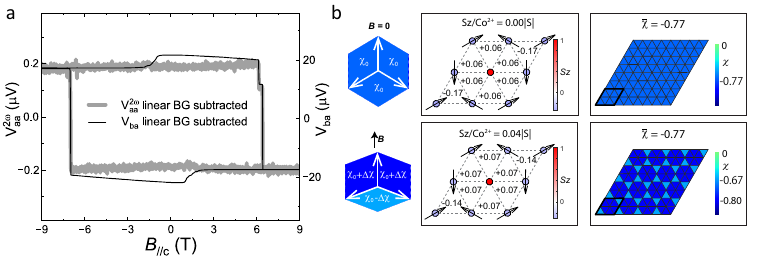}
    \caption{\textbf{a} The field dependence second-harmonic voltage and Hall voltage at 3 K, after subtraction of the linear background. The second-harmonic voltage remains robust across the measured field range, whereas Hall voltage captures the metamagnetic transition. \textbf{b} 2D analogue of the spin texture and the chirality defined by it, and calculated net magnetization and scalar spin chirality with and without an external magnetic field. In the spin model for net magnetization, the arrows represent the in-plane component of the spins and the color in circles represents the out-of-plane component of the spins. The local out-of-plane magnetization $S^{z}_{\mathrm{local}}$ and the local scalar chirality $\chi_{\mathrm{local}}$ are indicated by the values and colors on each triangular plaquette, respectively. The parameters used in the simulation are $J_1 = 1$, $J_2 = 0.75$, $D = 0.05$, $B = 0.25$ and $K = 0.01$. For simulations with an external field, we choose $H = 0.3$, corresponding to a similar energy scale as in the experiment, estimated using data from Ref.~\cite{Park2023}}
\end{figure}

Because of the translational symmetry of the spin texture, the total energy can be obtained by evaluating the four inequivalent spins within a single magnetic unit cell. In the energy minimization, the spin amplitudes are fixed, while the polar angles are treated as free variables. The azimuthal angles of the four spins are fixed to $0$, $\pi$, $-\pi/3$, and $\pi/3$, respectively, revealing the fundamental features of the scalar spin chirality.

At zero external field, the energy landscape exhibits two degenerate local minima. One corresponds to the spin configuration shown in the figure, while the other corresponds to its time-reversed counterpart. Upon applying a finite external field, the two minima become non-degenerate, resulting in an energetic preference for one of the two polarities.

To characterize the net magnetization and scalar spin chirality, we evaluate their local distributions on each triangular plaquette formed by three nearest-neighbor spins. The local scalar spin chirality is defined as
\begin{equation}
\chi_{\mathrm{local}} = \mathbf{S}_i \cdot \left( \mathbf{S}_j \times \mathbf{S}_k \right)
\end{equation}
The local net magnetization of each plaquette is defined as
\begin{equation}
S^{z}_{\mathrm{local}} = \frac{S_i^{z} + S_j^{z} + S_k^{z}}{6}
\end{equation}
We find that eight such plaquettes form the basic unit of the local chirality and magnetization distribution pattern. Since transport measurements probe the global state of the device, we calculate the average chirality over the eight plaquettes as
\begin{equation}
\bar{\chi} = \frac{1}{8} \sum \chi_{\mathrm{local}}
\end{equation}
The net magnetization per Co$^{2+}$ ion within this eight-plaquette unit is obtained as
\begin{equation}
 S^{z}/\mathrm{Co^{2+}} = \frac{1}{4} \sum S^{z}_{\mathrm{local}}
\end{equation}

The simulation shows that at zero field the four inequivalent spins of the magnetic unit cell in the purely chiral order sit at polar angles of $0^\circ$ and approximately $109^\circ$, close to an ideal tetrahedron, so that the net magnetization nearly cancels while the scalar chirality is uniform across all plaquettes. Applying a field along $c$ cants the three tilted spins upward and raises the net magnetization. The chirality of the plaquettes formed by three canted spins is reduced as the solid angle they subtend shrinks, but each of those spins is shared with neighbouring plaquettes that also contain the upright spin, on which the chirality is correspondingly enhanced. Averaged over the lattice these opposing changes compensate, so that the average scalar chirality remains constant to first order in the applied field while the magnetization grows. This plaquette locking explains both the flat plateaus of the SMD and its deviation in field from the anomalous Hall effect: the latter follows the rising magnetisation, the former tracks the field-independent chirality.

\bigskip

\section{Symmetry analysis of the second-order conductivity $\sigma_{ijk}$}

\begin{table}[h!]
\centering
\begin{tabular}{c l l l l l l l}
\toprule
$\sigma_{ijk} \rightarrow \sigma_{i\alpha}$ & \multicolumn{7}{l}{$jk \rightarrow \alpha$ — driving electric field} \\
\midrule
 &   & 1 & 2 & 3 & 4 & 5 & 6 \\
 & 1 & 0 & 0 & 0 & $\boldsymbol{\sigma_{14}}$ & 0 & 0 \\
$i$ — current & 2 & 0 & 0 & 0 & 0 & $-\sigma_{14}$ & 0 \\
 & 3 & 0 & 0 & 0 & 0 & 0 & 0 \\
\bottomrule
\end{tabular}
\caption{Nonrelativistic spin symmetry-allowed components of the second-order conductivity tensor in \CTS. Independent components are written in bold.}
\label{spingroupemcha}
\end{table}

\begin{table}[h!]
\centering
\begin{tabular}{c l l l l l l l}
\toprule
$\sigma_{ijk} \rightarrow \sigma_{i\alpha}$ & \multicolumn{7}{l}{$jk \rightarrow \alpha$ — driving electric field} \\
\midrule
 &   & 1 & 2 & 3 & 4 & 5 & 6 \\
 & 1 & $\boldsymbol{\sigma_{11}}$ & $-\sigma_{11}$ & 0 & $\boldsymbol{\sigma_{14}}$ & 0 & 0 \\
$i$ — current & 2 & 0 & 0 & 0 & 0 & $-\sigma_{14}$ & $-\sigma_{11}$ \\
 & 3 & 0 & 0 & 0 & 0 & 0 & 0 \\
\bottomrule
\end{tabular}
\caption{Relativistic magnetic symmetry-allowed components of the second-order conductivity tensor in \CTS. Independent components are written in bold.}
\label{maggroupemcha}
\end{table}

The crystal structure of \CTS~belongs to the $P6_{3}22$ space group (\#182). Within the spin group theory (when neglecting relativistic spin-orbit coupling), the non-coplanar magnetic ordering belongs to the $P{}^{3}6_{3}{}^{m}2{}^{m}2$ spin space group.
In the relativistic picture, the magnetic ordering breaks the six-fold rotation symmetry and the resulting magnetic space group is $P32'1$ (\#150.27).

In Table~\ref{spingroupemcha} and Table~\ref{maggroupemcha}, we list the symmetry-allowed components of the second-order conductivity tensor $\sigma_{ijk}$ (Jahn symbol $V[V2]$) within the spin and magnetic group of \CTS, obtained via \textsc{Stensor}~\cite{elcoro_automatic_2026,etxebarria_crystal_2025}. Since the tensor is symmetric in the last two dimensions ($\sigma_{ijk} = \sigma_{ikj}$), we use the shortened Voigt notation. Measured voltage component $V_{ij}^{2 \omega}$ corresponds to conductivity component $\sigma_{ijj}$.

Experimental results show a nonzero $V^{2\omega}_{aa} \sim \sigma_{111}$ component and zero $V^{2\omega}_{cc} \sim \sigma_{333}$ component. This is consistent with the relativistic form of the tensor shown in Table~\ref{maggroupemcha}. Remarkably, the $\sigma_{111}$ component is forbidden in the nonrelativistic form in Table~\ref{spingroupemcha}, implying that the measured $V_{2\omega}$ in \CTS~is of relativistic origin.

\section{Tight-binding model and nonlinear Drude response}

We consider a two-dimensional tight-binding model on a triangular
lattice with four magnetic sublattices, following the minimal
noncoplanar model of Ref.~\cite{mitscherling2025orbitalmagnetizationparalleltransport}. The Hamiltonian is 

\begin{equation}
\hat H =
-\sum_{\langle ij\rangle,\sigma}
t_{ij}\hat c_{i\sigma}^{\dagger}\hat c_{j\sigma}
-J\sum_{i,\sigma,\sigma'}
\hat c_{i\sigma}^{\dagger}
\left(
\bm S_i\cdot\bm\sigma_{\sigma\sigma'}
\right)
\hat c_{i\sigma'}
+i\lambda_R
\sum_{\langle ij\rangle}
\hat c_i^{\dagger}
\left[
(\hat{\bm z}\times\hat{\bm d}_{ij})
\cdot\bm\sigma
\right]
\hat c_j+\mathrm{H.c.},
\label{eq:TB_Rashba}
\end{equation}
where $t_{ij}$ denotes the spin-independent nearest-neighbor hopping,
$J$ is the local exchange coupling, $\bm S_i$ denotes the noncoplanar
magnetic moments, and $\hat{\bm d}_{ij}$ is the unit vector along the
bond $i\rightarrow j$. In the present calculations we use a distorted
but fully compensated tetrahedral magnetic structure, with the moment
amplitudes chosen according to the ratio $1.60/1.45$ while maintaining
$\sum_i\bm S_i=0$. Following the energy convention of the original
model, we set the nearest-neighbor hopping and exchange scale to unity;
in our implementation $t=-1$ and $J=1$. To phenomenologically model
the strong spin-orbit-coupling character of the material, motivated by
the substantial Ta contribution close to the Fermi level, we take
$\lambda_R=0.5$. The Rashba term represents an inversion-asymmetric
spin-dependent hopping with an effective spin-orbit field proportional
to $\hat{\bm z}\times\hat{\bm d}_{ij}$, whereas the ordinary hopping
$t_{ij}$ remains spin independent. Reversal of the magnetic domain is
implemented as $\bm S_i\rightarrow-\bm S_i$, while the Rashba coupling
is kept unchanged because it originates from the structural inversion
asymmetry.

The nonlinear longitudinal response is evaluated from the resulting
band structure using \textsc{WannierBerri} \cite{wannierberritsirkin2021high}. Within the constant
relaxation-time approximation, the intraband nonlinear Drude
conductivity is
\begin{equation}
\sigma^{\mathrm{NLDrude}}_{abc}
=
-\frac{e^3\tau^2}{\hbar^3}
\sum_n
\int_{\mathrm{BZ}}
\frac{d^d k}{(2\pi)^d}\,
v_{n,a}v_{n,b}v_{n,c}
\frac{\partial^2 f_0}{\partial\varepsilon_n^2},
\qquad
v_{n,a}
=
\frac{\partial\varepsilon_n}{\partial k_a},
\label{eq:NLDrude}
\end{equation}
where $f_0$ is the equilibrium Fermi--Dirac distribution and we use
a relaxation time $\tau=10\,\mathrm{fs}$. The Bloch Hamiltonian in
Eq.~(\ref{eq:TB_Rashba}) is supplied directly to
\textsc{WannierBerri}, and the Brillouin-zone integral is evaluated on
a dense two-dimensional momentum mesh while varying the chemical
potential. Since the tight-binding model represents an effectively
two-dimensional conducting layer, the calculated sheet response is
converted to an effective three-dimensional conductivity through
\begin{equation}
\sigma^{3\mathrm{D}}_{abc}
=
\frac{\sigma^{2\mathrm{D}}_{abc}}{d},
\qquad
d=5\,\text{\AA},
\label{eq:2D3D}
\end{equation}
where the effective thickness is chosen to be of the order of one half
of the experimental out-of-plane lattice parameter. For the two
spin-reversed magnetic configurations,
$\varepsilon_{n,-\bm S}(\bm k)
=
\varepsilon_{n,+\bm S}(-\bm k)$,
and the corresponding nonlinear Drude response changes sign,
\begin{equation}
\sigma^{\mathrm{NLDrude}}_{abc}[-\bm S]
=
-\sigma^{\mathrm{NLDrude}}_{abc}[+\bm S].
\label{eq:NLDrude_reverse}
\end{equation}

\end{document}